\documentclass[final,5p,times,twocolumn]{elsarticle}
\usepackage{color,amssymb}

\usepackage{hyperref}

\journal{Physics Letters B}

\begin{document}

\begin{frontmatter}
\title{Testing universal dark-matter caustic rings with galactic rotation curves}

\author[MSU,INR]{Daniil Davydov}
\ead{davydov.dd19@physics.msu.ru}

\affiliation[MSU]{organization={Physics Department, M.V.~Lomonosov Moscow State University},
            addressline={1-2 Leninskie Gory}, 
            postcode={119991}, 
            city={Moscow},
            country={Russia}}

\affiliation[INR]{organization={Institute for Nuclear Research of the Russian Academy of Sciences},
            addressline={60th October Anniversary Prospect 7A}, 
            postcode={117312}, 
            state={Moscow},
            country={Russia}}
\author[INR,MSU]{Sergey Troitsky}

\date{\rm INR--TH-2022-012}
\begin{abstract}
Infall of cold dark matter on a galaxy may result in caustic rings where the particle density is enhanced. They may be searched for as features in the galactic rotation curves. Previous studies suggested the evidence for these caustic rings with universal, that is common for different galaxies, parameters. Here we test this hypothesis with a large independent set of rotation curves by means of an improved statistical method. No evidence for universal caustic rings is found in the new analysis.
\end{abstract}




\end{frontmatter}

\section{Introduction}
\label{sec:intro}

The origin of the dark matter, invisible substance not explained in terms of the Standard Model of particle physics, is one of main unsolved problems in modern physics, see e.g.\ Ref.~\cite{BertoneTait} and references therein. Long ago, it was pointed out that in the process of accumulating of cold dark matter in galaxies, infall caustics may form, which are concentric structures of enhanced dark-matter density \cite{SikivieIpser1992,STW1,STW2,DuffySikivie-MW2008,Dolag:2012uq}. Inner caustics are ring-like structures \cite{NatarajanSikivie:inner=ring} which were suggested to be typical for the case of the axion dark matter \cite{Sikivie2010-AX}.

These overdensities should reveal themselves as features in the rotation curves of galaxies. However, on average, rotation curves are smooth and universal, see e.g.\ Refs.\ \cite{Salucci1,Salucci2}. The features caused by caustics are weak enough, so it is a difficult task to distinguish them from variations of observed velocities caused by local streams and even from statistical fluctuations. Positions of the caustics are governed by the distribution of angular momenta of infalling dark-matter particles. Based on the observations of such bumps in the rotation curves of NGC~3198 \cite{Sikivie1997-NGC3198} and the Milky Way \cite{Sikivie-MilkyWay2001}, Kinney and Sikivie proposed \cite{KinneySikivieCaustics} that this distribution is universal and therefore positions of the caustic rings in different galaxies should coincide up to a well-determined rescaling. This opened a possibility to use ensembles of galaxies to search for coinciding bumps in their (rescaled) rotation curves by means of stacking, which is expected to suppress features caused by local streams and fluctuations but enhances possible universal bumps associated with caustics. Moreover, the authors of Ref.~\cite{KinneySikivieCaustics} analyzed a set of 32 high-quality galactic rotation curves available by that time \cite{1991,1996} and found indications to the existence of two caustic-associated bumps, with the reported statistical significance between 2.4$\sigma$ and 3.0$\sigma$. 

To our best knowledge, these results have never been tested with newer data, though a number of rotation curves of similar quality have become available. Ref.~\cite{Dumas:MWcaustics} considered the caustic model for the Milky Way in a different approach. Refs.~\cite{deBoer1011,Huang1604,Chakrabarty2007} looked for features in the Milky-Way rotation curves, while Ref.~\cite{Onemli0710} claimed an observation of a caustic ring in a particular galaxy clusters. Therefore, a statistical analysis of possible universal caustic rings was performed only in Ref.~\cite{KinneySikivieCaustics}. The purpose of the present work is to test the same hypothesis of universal caustic rings \cite{KinneySikivieCaustics} with a large independent set of rotation curves published since then. 

The rest of the paper is organised as follows. In Section~\ref{sec:data} we describe the data on galactic rotation curves we use. Section~\ref{sec:anal} presents the analysis procedure. For each galaxy, we closely repeat the procedure of Ref.~\cite{KinneySikivieCaustics}, which we describe for convenience in Section~\ref{sec:anal:individ}. The statistical analysis of the ensemble of rotation curves is discussed in Section~\ref{sec:anal:ensemble}. In Section~\ref{sec:res}, we present and discuss our results. The main test of the hypothesis of universal caustic rings is performed with a set of rotation curves independent from that used in Ref.~\cite{KinneySikivieCaustics}; its results are presented in Section~\ref{sec:res:indep}. In section~\ref{sec:res:oldsamples}, we show how these results are affected by inclusion of the galaxies from the original sample of Ref.~\cite{KinneySikivieCaustics}. We discuss differences with Ref.~\cite{KinneySikivieCaustics} in our methods and results in Section~\ref{sec:res:discussion} and briefly conclude in Section~\ref{sec:concl}.

\section{Data on rotation curves}
\label{sec:data}
We use the \emph{Spitzer Photometry and Accurate Rotation Curves} (SPARC) database of rotation curves\footnote{Data available from \url{http://astroweb.cwru.edu/SPARC}.} for 175 galaxies with various morphologies and luminosities \cite{2016}. These rotation curves were carefully selected from a number of previous studies, as described in Ref.~\cite{2016}. The rotation velocities were obtained from interferometric observations of atomic hydrogen which is one of the best tracers of kinematics of nearby galaxies for a wide range of radial distances, which is important for our purposes.

The previous study \cite{KinneySikivieCaustics} made use of a sample of 32 rotation curves taken from Refs.~\cite{1991,1996}. Of these 32 galaxies, 29 are present in the SPARC database. In order to perform an independent test of the hypothesis of Ref.~\cite{KinneySikivieCaustics}, we remove these 29 rotation curves from the sample. In addition, 25 curves have been removed to satisfy the quality condition described below in Sec.~\ref{sec:anal:individ}. The main, independent from Ref.~\cite{KinneySikivieCaustics}, sample we use thus contains 121 rotation curves.

In Sec.~\ref{sec:res:discussion}, we compare our results with those of Ref.~\cite{KinneySikivieCaustics}. There, we use also the original sample of 32 rotation curves from Refs.~\cite{1991,1996}.

\section{Data analysis}
\label{sec:anal}
\subsection{Individual rotation curves}
\label{sec:anal:individ}
We are looking for caustic rings, where the concentration of dark matter is enhanced. In these areas, the rotation velocity should increase compared to other regions, resulting in features in the rotation curve. However, even without caustics, the rotation curves are not flat, and individual gas clouds observed in a galaxy may have peculiar velocities independent of any caustic structure. To quantify the presence or lack of features, the authors of Ref.~\cite{KinneySikivieCaustics} proposed to use the procedure for separation the background rotation of the galaxy and for quantifying the ``noise''. The procedure, which we follow exactly here, is as follows.

Each rotation curve is a set of measurements \((r_{i}, v_{i}\pm\Delta v_i)\), where $r_i$ is the radial distance, $v_i$ is the velocity and $\Delta v_i$ is the uncertainty of the velocity. Assume for the moment that the rotation curve at large radial distances is flat and denote as  \(v_{\rm rot}\)  the velocity in the part where it reaches the plateau. Then the hypothesis of Refs.~\cite{Sikivie1997-NGC3198,KinneySikivieCaustics} predicts caustic rings at
$$
\tilde r= a_n
\frac{j_{\rm max}}{0.25} \frac{0.7}{h},
$$
where the rescaled radius is defined as
\begin{equation}
\label {eq:1}
\tilde{r} \equiv r \left(\frac{220~\mbox{km/s}}{v_{\rm rot}}\right),
\end{equation}
$n=1,2,3 \dots$ enumerates the caustic rings formed by the particles experiencing the $n$th infall,
$$a_n=\left\{ 39,19.5,13,10,8,\dots\right\}~\mbox{kpc}$$
are the universal rescaled positions of these caustics, 
$j_{\rm max}$ is the peak of the distribution of dimensionless angular momenta of dark-matter particles and $h$ is the dimensionless local Hubble constant. The universality hypothesis of Ref.~\cite{KinneySikivieCaustics} is that for all galaxies, $j_{max}=0.27$ and therefore the rotation-curve bumps of various galaxies should coincide in terms of $\tilde r$, Eq.~(\ref{eq:1}). We note that there is an uncertainty related to the value of $h$ used in the distance determination. While Ref.~\cite{KinneySikivieCaustics} suggests to use the value which observers used in constructing the rotation curve, the Hubble distances are not precise for nearby galaxies in the sample and other methods of distance determination are used whenever possible. For 97 of 175 galaxies in the SPARC sample, $h=0.73$ was assumed; other 78, non-Hubble, distances correspond to this value of $h$ \cite{2016}. In the present work, we do not stick to particular values of $a_n$, $j_{\rm max}$ and $h$ and search for any rotation-curve bumps at coinciding $\tilde r$, thus testing the hypothesis of the universal $j_{\rm max}$ without assuming its particular value.

In practice, the rotation curves are never exactly flat, and the plateau regions, if any, are reached at various radial distances. Ref.~\cite{KinneySikivieCaustics} suggested to use the average rotation velocity in the outer part of the galaxy as $v_{\rm rot}$. To calculate it, the data points are fitted by a smooth curve (a line or a polynomial was used in Ref.~\cite{KinneySikivieCaustics}). It is assumed that the approximately flat part of the rotation curve corresponds to the rescaled radii \(\tilde{r_{i}} \gtrsim  10~\mbox{kpc} \), and only this range contributes to the average. In our work, we follow the same prescription and determine the fitting parabola $\bar v(\tilde r) = c_0+c_1 \tilde r + c_2 \tilde r^2$, then the average value $v_{\rm rot}= \langle v(\tilde r) \rangle_{\tilde r\ge 10~\rm{kpc}}$ is used in Eq.~(\ref{eq:1}). Like it was done in Ref.~\cite{KinneySikivieCaustics}, we tested polynomials of various degrees in $\tilde r$ as the fitting functions for $\bar v(\tilde r)$ and, like in Ref.~\cite{KinneySikivieCaustics}, the results were not changed.

Like in Ref.~\cite{KinneySikivieCaustics}, we need a little complication of the procedure because the definition of $\tilde r$, and hence the condition \(\tilde{r_{i}} \ge 10~\mbox{kpc} \), depend on $v_{\rm rot}$. The rescaled rotation curve is determined iteratively. Initial data are fitted by the parabola, \(\bar{v_0}({r})\), and the mean $v_{\rm rot}$ is calculated for this parabola over $r_i \ge 10$~kpc. Then Eq.~(\ref{eq:1}) is used to obtain rescaled radii $\tilde r$, and all points with  
\(\tilde{r_{i}} 
< 10\)~kpc are removed. At the next iteration, the new fitting function \(\bar{v_{1}}({\tilde r})\) is constructed, new $v_{\rm rot}$ and $\tilde r_i$ are calculated and the condition $\tilde r_i \ge 10$~kpc is checked again. This iteration procedure converges fastly, normally in one or two iterations. As a result, we obtain a set of rescaled data points, (\(\tilde r_{i}, v_{i} \pm \Delta v_{i})\), where all \(\tilde r_i\ge 10\)~kpc, and the corresponding fitting parabola $\bar v(\tilde r)$. Note that the parabolic fits always give a good description of the outer parts of the rotation curves.

At the next step, we will be interested in bumps, that is deviations from the smooth rescaled rotation curve $\bar v(\tilde r)$. Therefore, the output of this step for each galaxy is a set of $(\tilde r_i,\sigma_i)$, where we defined $\sigma_i\equiv \left( v_i-\bar v(\tilde r_i)\right)/\Delta v_i$. Note that since the parabola $\bar v(\tilde r)$ for three data points is determined precisely, we require at least four data points to search for bumps against the background of this parabola. This quality cut removes 25 galaxies from the main data set (and none from the set used in Sec.~\ref{sec:res:oldsamples}).

This procedure, similar to the one which resulted in the positive signal in Ref.~\cite{KinneySikivieCaustics}, may be considered as a part of the hypothesis to be tested, so we keep it in the present study.

\subsection{Ensemble of rotation curves}
\label{sec:anal:ensemble}
To search for features at the same rescaled radii $\tilde r$ present in rotation curves of different galaxies, we adopt the unbinned likelihood method, in which a continuous function $L(\tilde r)$ is constructed from the entire data set of all galaxies. To estimate the statistical significance of any potential feature, including a universal caustic ring, we use the Monte-Carlo method with the same function $L$ constructed from simulated data, as described in Sec.~\ref{sec:anal:significance} below.

The statistical method we use for the ensemble of rotation curves is different from the one used in Ref.~\cite{KinneySikivieCaustics}. The reasons for this change are discussed, and comparison with the method of Ref.~\cite{KinneySikivieCaustics} is presented in Sec.~\ref{sec:res:discussion}.

For each galaxy, we start with the set \((\tilde r_{i}, \, \sigma_{i})\) obtained as described in Sec.~\ref{sec:anal:individ}. For each \(\sigma_{i}\) we assign a \(p_{i}\) value \((0 \leqslant p_{i} \leqslant 1 )\) with the meaning of the probability of a random deviation to $\ge \sigma_i$ for the Gaussian distribution with the mean \(0\) and variance \(1\),
\begin{equation}
p_{i} =
1-\mbox{CDF}(\sigma_i)=\frac{1}{2}\left(1-\mbox{erf}(\frac{\sigma_i}{\sqrt{2}})\right).
\end{equation}
For each galaxy $j$, we then obtain a function \(p_j(\tilde{r})\) by the linear interpolation of the corresponding points \((\tilde{r_{i}}, \, p_{i})\) for this galaxy. Note that \(p_j(\tilde{r})\) is determined on an interval between the minimal and maximal values of $\tilde r_i$, and these values are different for different galaxies. 

We are now ready to construct the averaged likelihood function \(L(\tilde{r})\) as
\begin{equation}
L(\tilde{r}) = - \sum\limits_{j=1}^{N(\tilde{r})} \frac{\log [p_j
(\tilde{r})]}{N(\tilde{r})},
\end{equation}
where \(N(\tilde{r})\) is the number of galaxies, for which $p_j$ is determined at the point \(\tilde{r}\). 

The interpretation of this likelihood function is straightforward. Each galaxy may have outlaying points in the rotation curve, which may or may not be associated with caustics. Random statistical fluctuations and individual rotation-curve features caused by particular velocity flows are expected to occur at different $\tilde r$ and therefore to cancel in the sum $L$. Contrary, coinciding bumps related to universal caustics would add coherently resulting in a peak in $L(\tilde r)$.

Note that details of the construction of the likelihood function, e.g.\ the use of the Gaussian $p_i$ or the choice of the interpolation method, are not important since we never interpret the actual value of $L$ in terms of the significance, using instead the Monte-Carlo approach described below.

\subsection{Monte-Carlo estimate of significance}
\label{sec:anal:significance}
To estimate the significance of potential universal features in the rescaled rotation curves, we follow the standard parametric bootstrap approach, see e.g.\ Ref.~\cite{NR}. We use the maximal value $L_{\rm max}$ of $L(\tilde r)$ over the interval of interest, 10~kpc~$\le \tilde r \le$~75~kpc, which covers well the first two caustic rings discussed in Ref.~\cite{KinneySikivieCaustics}. The lower limit of the interval is implied by the procedure used to construct $\sigma$ while the upper one is determined by the lack of statistics of rotation-curve measurements at large radial distances. 

To simulate artificial rotation curves without universal features, we assume that the fitted functions $\bar v(\tilde r)$ represent the true smooth rotation curves, the measurements were done at the same sets of $r_i$ as in the real data, and the measurement errors are Gaussian with the same widths as quoted for the real data. Within these assumptions, we obtain a simulated data set and process it in the same way as the real data, see Sec.~\ref{sec:anal:ensemble}. 

We repeat $M$ times the same procedure with simulated rotation-curve measurements and obtain a set of simulated $L_{\rm max}^{(k)}$, $k=1,\dots,M$. The significance of the strongest universal feature in the set of rescaled rotation curves is determined by the p-value, counting how often the observed or larger value of $L_{\rm max}$ happens by chance in the simulated sets, that is the number of cases for which $L_{\rm max}^{(k)}>L_{\rm max}$ divided by $M$.

\section{Results and discussion}
\label{sec:res}
\subsection{Results for the main sample}
\label{sec:res:indep}
\begin{figure}
\center\includegraphics[width=0.95\columnwidth]{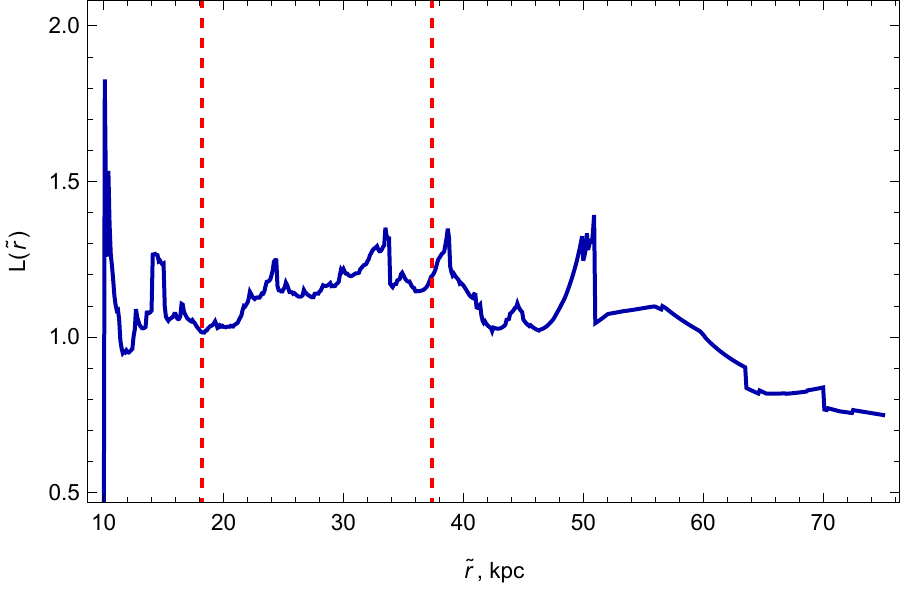}
\caption{The function \(L(\tilde{r})\) for the main sample of 121 galaxies which does not include those studied in Ref.~\cite{KinneySikivieCaustics}. The vertical dashed lines indicate the expected positions of the $n=1,2$ caustic rings claimed in Ref.~\cite{KinneySikivieCaustics}, corrected for the slightly different value of $h$.}
\label{fig:146}
\end{figure}

\begin{figure}
\center\includegraphics[width=0.95\columnwidth]{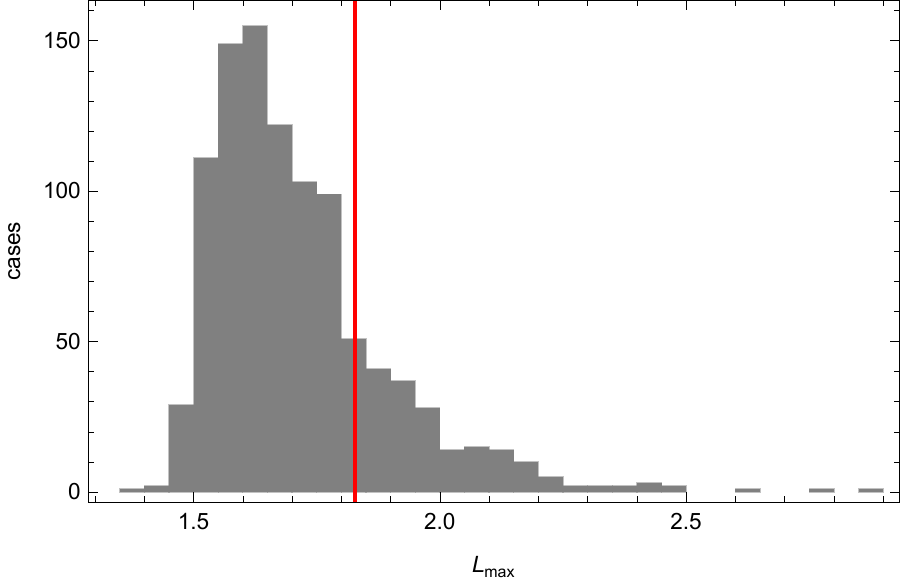}
\caption{The distribution of the maxima of 1000 Monte-Carlo simulated $L(\tilde r)$ functions for the main sample of rotation curves. The vertical line indicates the value of $L_{\rm max}$ obtained for the data.}
\label{fig:168-22}
\end{figure}

The application of the procedure described in Sec.~\ref{sec:anal:individ}, \ref{sec:anal:ensemble} to the main data set of 121 rotation curves not used in previous search for caustics, described in Sec.~\ref{sec:data}, results in the $L(\tilde r)$ function presented in Fig.~\ref{fig:146}.
No significant peaks of \(L(\tilde{r})\) are observed at the expected positions of caustic rings found in Ref.~\cite{KinneySikivieCaustics}, nor elsewhere. Indeed, 
Figure~\ref{fig:168-22} 
presents the distribution of $L_{\rm max}^{(k)}$ for $k=1,\dots, M=1000$ Monte-Carlo simulated data sets assuming no universal caustics and only statistical fluctuations in the rotation-curve measurements. The observed $L_{\rm max}\approx 1.83$ or larger was found 202 times out of 1000, resulting in the p-value of 0.2 for the null hypothesis of the absence of universal caustic rings. This confirms that the smooth, bumpless $\bar{v}(\tilde r)$ quadratic functions we used describe well the rotation curves.
\begin{figure*}
\begin{center}
\includegraphics[width=0.95\columnwidth]{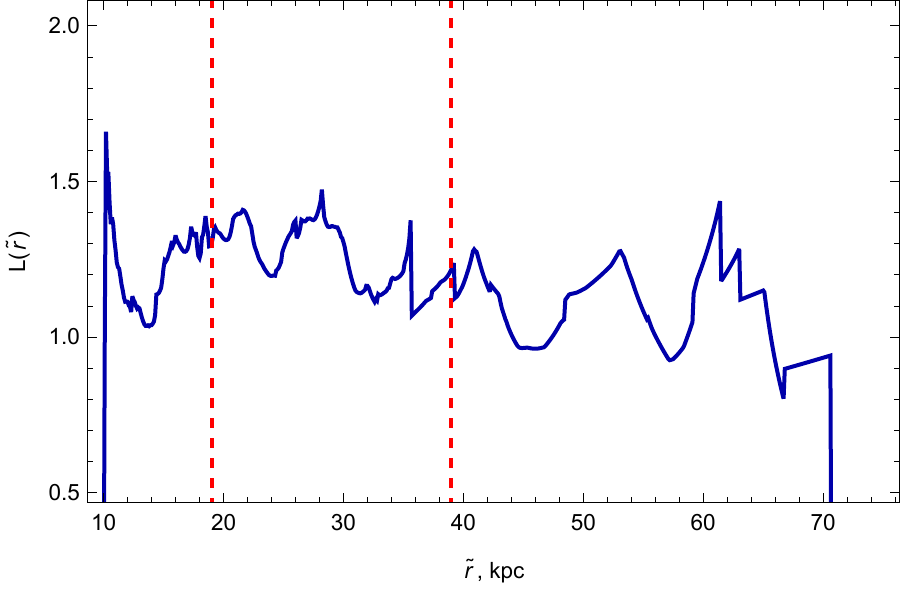}~~~%
\includegraphics[width=0.95\columnwidth]{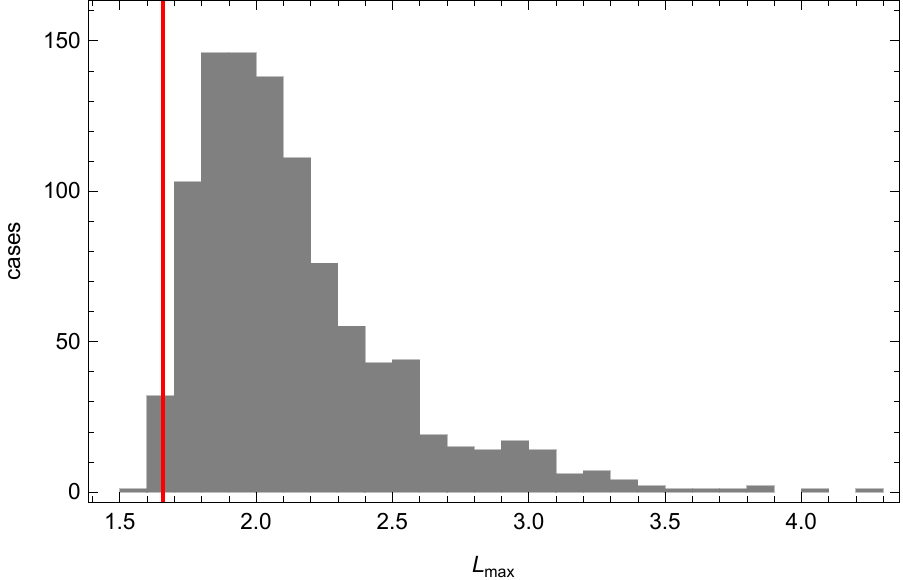}
\end{center}
\caption{Results for the sample of 32 rotation curves studied in Ref.~\cite{KinneySikivieCaustics}. Left: the function \(L(\tilde{r})\). The vertical dashed lines indicate the expected positions of the $n=1,2$ caustic rings claimed in Ref.~\cite{KinneySikivieCaustics}. Right: the distribution of the maxima of 1000 Monte-Carlo simulated $L(\tilde r)$. The vertical line indicates the value of $L_{\rm max}$ obtained for the data.}
\label{fig:32G}
\end{figure*}
\begin{figure*}
\begin{center}
\includegraphics[width=0.95\columnwidth]{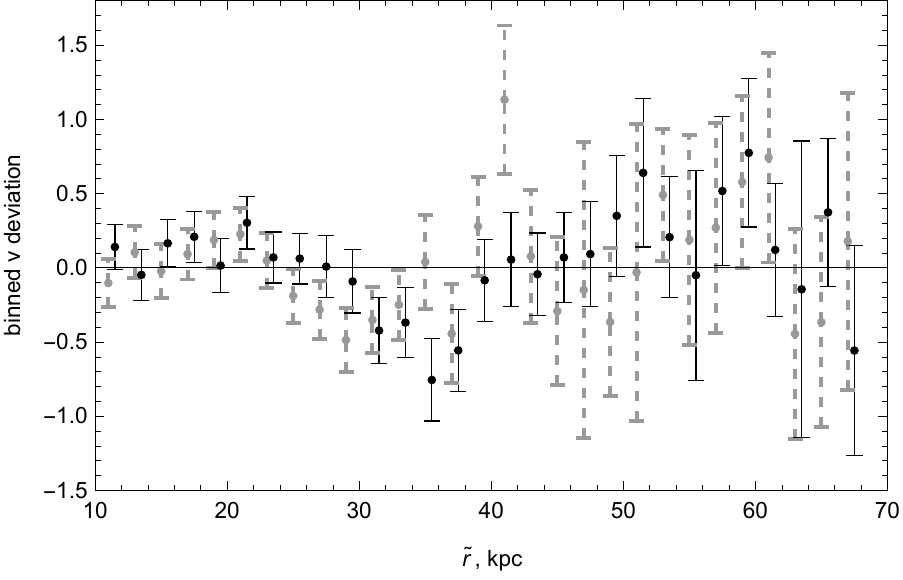}~~~%
\includegraphics[width=0.95\columnwidth]{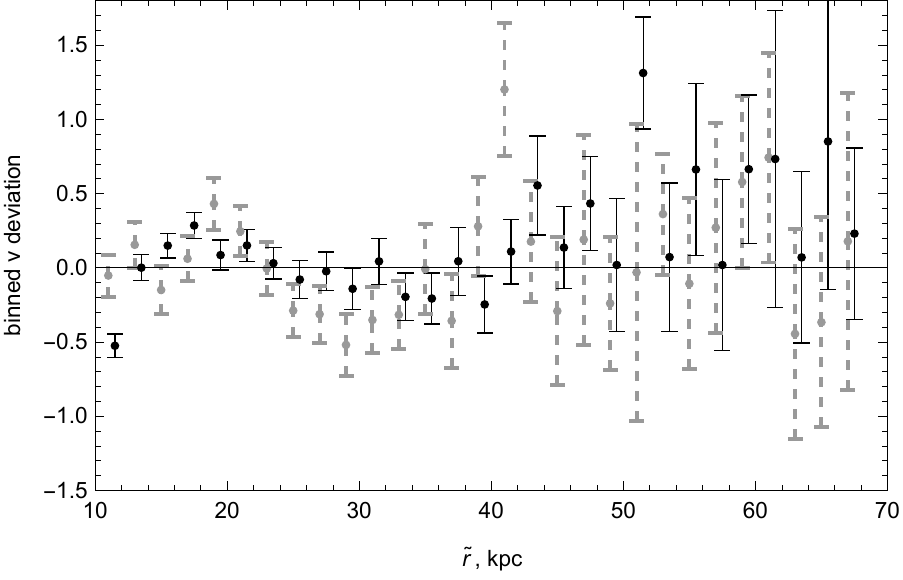}
\end{center}
\caption{Comparison of the new results with those of Ref.~\cite{KinneySikivieCaustics} by the binned method of Ref.~\cite{KinneySikivieCaustics}, see the text for details. Grey points with dashed error bars: the rotation curves used in Ref.~\cite{KinneySikivieCaustics}; black points with full error bars: new rotation curves from the SPARC database. Left panel: 29 galaxies present in both samples. Right panel: 32 galaxies in the old sample and 121 other galaxies in the new sample.}
\label{fig:sikplots}
\end{figure*}

\subsection{Results for the previously used sample}
\label{sec:res:oldsamples}
The results obtained above for the statistically independent sample of rotation curves do not support the indications to the presence of universal caustic rings claimed in Ref.~\cite{KinneySikivieCaustics}. It is therefore instructive to repeat our analysis for the original sample used in that study. Figure~\ref{fig:32G}
presents the results of this analysis. Interestingly, it also does not reveal any significant excesses at the expected positions of the universal caustics. Contrary, we find the p-value of 0.98 for the null hypothesis, which means that the consistence between data and expectations is ``too good'', which may happen for overestimated statistical errors in the initial data. 

\subsection{Discussion: comparison with the previous result}
\label{sec:res:discussion}
There are two main differences between the statistical analysis used in Ref.~\cite{KinneySikivieCaustics} and that of the present work. Firstly, we use the unbinned likelihood while the analysis of Ref.~\cite{KinneySikivieCaustics} was based on binning. Secondly, we use Monte-Carlo simulations based on the null hypothesis to estimate the significance, while Ref.~\cite{KinneySikivieCaustics} assigned statistical errors to the binned data by hand. We note that the resulting significance reported in Ref.~\cite{KinneySikivieCaustics} does not account for the selection of the bin with the strongest deviation (the look-elsewhere effect), nor for the choice of the bin size. For the 30 bins, Ref.~\cite{KinneySikivieCaustics}  reports a $2.6\sigma$ significance, that is the p-value of 0.01; since the data in the bins are independent, a reasonable estimate of the look-elsewhere correction would be to multiply the p-value by the number of bins, which would result in the post-trial significance fully consistent with the null hypothesis. Ref.~\cite{KinneySikivieCaustics}  also considers the case of 60 bins and obtains $3.0\sigma$ then; multiplied by the number of bins and by the number of bin-size trials, that is 2, this pre-trial p-value gives the same result for the post-trial significance. 

For completeness of the comparison, we present here also the results of the analysis of the new data with exactly the same procedure as the one used in Ref.~\cite{KinneySikivieCaustics}. The main result of that approach is presented as a plot with the normalized derivations of observed velocities from the fitted ones, averaged over bins in $\tilde r$, cf.\ Fig.~1 of Ref.~\cite{KinneySikivieCaustics}. The normalization used for each measurement is the dispersion of the data for a particular rotation curve, not the individual measurement error, see Ref.~\cite{KinneySikivieCaustics} for details. The error bars assigned to each bin were $1/\sqrt{N_i}$, where $N_i$ is the number of measurements in the bin. Note that we do not find this procedure optimal for the reasons discussed above and never use it to obtain our results, making use of the unbinned likelihood and Monte-Carlo simulations instead. Exclusively for illustration, we however present the corresponding plots in Fig.~\ref{fig:sikplots}. Its left panel compares the data for 29 galaxies present in both samples, that of Ref.~\cite{KinneySikivieCaustics} and SPARC. One can see that the bumps claimed in Ref.~\cite{KinneySikivieCaustics} disappeared in the new data, consistent with being caused by statistical fluctuations. The right panel of Fig.~\ref{fig:sikplots} compares the data for all 32 rotation curves \cite{KinneySikivieCaustics} with the independent set of 121 SPARC rotation curves we use in Sec.~\ref{sec:res:indep}. The result again does not confirm the claims of Ref.~\cite{KinneySikivieCaustics}, even if their method is used.

Therefore, we believe that the universal caustics observed in Ref.~\cite{KinneySikivieCaustics} were in fact caused by statistical fluctuations, not seen in the new data even for the same galaxies. The significance of the features found there was overestimated because the look-elsewhere correction was not addressed.

\section{Conclusions}
\label{sec:concl}
In the present paper, the hypothesis of the universality of caustic rings in galaxies, proposed in Ref.~\cite{KinneySikivieCaustics}, is tested. We use the SPARC sample of galactic rotation curves \cite{2016} from which we remove the galaxies used in the original study \cite{KinneySikivieCaustics}. After additional quality cuts, the main sample contains 121 rotation curves. We apply the unbinned likelihood method and estimate the significance of possible universal features in the rotation curves by means of the Monte-Carlo simulations. We find no indication of the presence of universal caustic rings: the data agrees with the expectations of the null hypothesis with the p-value of 0.20. We discuss the reasons for the discrepancy with the original proposal and find that the most probable reason is the use in Ref.~\cite{KinneySikivieCaustics} of the statistical method which overestimates the significance. We demonstrate in addition that the new data do not support the conclusions of Ref.~\cite{KinneySikivieCaustics} even if their method is used.

However, the stacking method, used in Ref.~\cite{KinneySikivieCaustics} and in the present work, tests only the universality of the caustics. The locations of the caustic rings may vary from one galaxy to another even in terms of the rescaled coordinate $\tilde r$, indicating a scatter in the angular momentum distributions of infalling dark matter. To find or exclude such non-universal caustics would be a much more difficult observational task because the effect of the matter overdensity on the rotation curve should be disentangled from local variations in the velocity field of the galactic gas. This may however be feasible in observations of two-dimensional velocity fields, see e.g.\ Refs.~\cite{THINGS,FabryPerotOld,FabryPerotNew}.
Observation or constraining of dark-matter caustics would shed light on the dynamical history of galaxies, the nature of the dark matter and its expected density near the Earth, being therefore important for both direct and indirect dark-matter detection \cite{indirect2005,indirect2007,direct2007,direct2020}.


We are indebted to D.~Levkov, A.~Plavin, K.~Postnov, G.~Rubtsov, I.~Tkachev and A.~Zasov for interesting and helpful discussions of various topics related to this work. DD thanks N.~Alutis for his help and for intense discussions of the analysis. ST thanks A.~Gribachev and T.~Glukhikh for collaboration at the initial stages of this work.
This work is supported by the Russian Science Foundation, grant 22--12--00215 (DD).


\bibliographystyle{elsarticle-num}
\bibliography{caustics}


\end{document}